\documentclass[pre,twocolumn,showpacs,floatfix,nofootinbib,groupaddress,showkeys]{revtex4}
\usepackage{graphicx}
\usepackage{amssymb}
\usepackage{amsmath}
\usepackage{bm}

\begin{document}

\title{Competition and cooperation in one-dimensional stepping stone models}
\author{K.~S.~Korolev}
\affiliation{Department of Physics, Massachusetts Institute of Technology, Cambridge, Massachusetts 02139, USA}
\author{David R. Nelson}
\affiliation{Department of Physics and FAS Center for Systems Biology, Harvard University, Cambridge, Massachusetts 02138, USA}
\begin{abstract}
Cooperative mutualism is a major force driving evolution and sustaining ecosystems. Although the importance of spatial degrees of freedom and number fluctuations is well-known, their effects on mutualism are not fully understood. With range expansions of microbes in mind, we show that, even when mutualism confers a distinct selective advantage, it persists only in populations with high density and frequent migrations. When these parameters are reduced, mutualism is generically lost via a directed percolation process, with a phase diagram strongly influenced by an exceptional DP2 transition.   
\end{abstract}

\date{\today}
\maketitle


Cooperation is at the heart of many complex systems~\cite{nowak:book,smith:games}. On an organism level, gut bacteria help their hosts digest cellulose. On an ecosystem level, plants often rely on fungi to receive important nutrients. Even human societies are products of cooperation between individuals. Despite the apparent advantage and pervasiveness of mutualistic interactions, their existence is often difficult to explain by a naive application of Darwinian natural selection: Cooperation can succumb to cheating~\cite{nowak:book} and, as we show here, to number fluctuations.    

To model complex interactions between individuals or species, Maynard Smith developed evolutionary game theory~\cite{smith:games}. The central idea of game theory is that the fitness of an organism depends on the frequency and types of encounters with other organisms in the population. Evolutionary games are usually analyzed using mean-field-type approximations, which neglect both spatial correlations and number fluctuations. However, these simplifications are not appropriate for natural populations living in spatially extended habitats and can miss important stochastic aspects of population dynamics. In particular, the interplay of stochasticity and spatial degrees of freedom leads to spatial demixing of different species or genotypes in the population~\cite{HallatschekNelson:ExperimentalSegregation,korolev:review}, which can significantly decrease the probability of mutualistic interactions. 

Following pioneering work of Nowak and May~\cite{nowak:spatialPD92}, several studies have investigated the effects of space on evolutionary games~\cite{hauert:snowdrift,roca:spatial_cooperation}~(and references therein) using simulations on a two-dimensional lattice with a single nonmotile individual per site.  Although these studies underscored the significant effects of spatial structure on evolutionary dynamics, outstanding issues remain. First, the outcomes of these lattice simulations are very sensitive to the exact rules of birth and death updates and interaction pattern between nearest neighbors~\cite{roca:spatial_cooperation}; as a result, these studies do not smoothly connect with the well-understood dynamics in spatially homogeneous~(well-mixed) populations. Moreover, it is not clear whether a model with a single nonmotile organism per site and nearest neighbor interactions is a good description of any species. Second, such models do not allow systematic investigation of the role of migration and the magnitude of number fluctuations, which are important for the applications of the theory to natural and experimental populations. Third, closely related voter models in two spatial dimensions have very slow logarithmic coarsening~\cite{Scheucher:SpinodalDecomposition}; 2d~simulations typically do not explore the time scales on which spatial demixing of species becomes important.

This letter studies a stepping stone model of one-dimensional population genetics~\cite{KimuraWeiss:SSM}. The stepping stone model preserves the interaction pattern of well-mixed populations, but includes migrations as well as number fluctuations, which are controlled by the population density. We focus on a one-dimensional model because stochastic effects are more pronounced in lower spatial dimensions~\cite{korolev:review}. More importantly, the spread of mutualism in two dimensions often occurs via a traveling reaction-diffusion wave, where the most important dynamics often occurs at a moving quasi-one-dimensional frontier~\cite{korolev:review}. We find that, for one-dimensional populations, mutualism persists in a much smaller region of parameter space than for well-mixed populations. Mutualism is particularly unstable against spatial demixing when the benefits to the interactants are unequal. The critical strength of mutualism required to sustain cooperation increases with migration rate and population density. As the strength of mutualism is reduced, the population undergoes a nonequilibrium phase transition in the universality class of either directed percolation~(DP) or~$\mathcal{Z}_{2}$ symmetric directed percolation~(DP2); see Ref.~\cite{hinrichsen:review} for a comprehensive review of DP models.

The stepping stone model~\cite{KimuraWeiss:SSM} we use to simulate the evolution of mutualism on a computer consists of demes~(islands) arranged on a line, with spacing~$a$. Each deme has~$N$ organisms, which can reproduce and migrate. One reproduction and one migration update in each deme constitute a time step. One generation, in this Moran process~\cite{korolev:review}, corresponds to~$N$ time steps because every individual in a deme is updated once on average. Organisms can migrate to one of the two nearest neighbors with equal probability, and each organism has a probability~$m$ to migrate during a generation time~$\tau$. Reproduction occurs within a deme by selecting a random individual to die and another individual to reproduce. The probability to be selected for reproduction is proportional to individual's fitness. To study mutualism, we assume that the fitness is a sum of two contributions: a background reproduction rate, which we scale to one for all organisms in the population, and a benefit due to mutualistic interactions with other organisms in a deme (e.g. due to exchanging nutrients). Let the benefit to the organism of type~$i$ from interacting with the organism of type~$j$ be~$a_{ij}$. If the types fractions within a deme are~$f_{i}$, then the corresponding fitnesses~$w_{i}$ in a given generation are~$w_{i}=1+\sum_{j}a_{ij}f_{j}$ because the increases in growth rate due to mutualism should be weighted by the density of cooperating organisms. 

For simplicity, we consider only two cooperating species~(or genotypes) and let the frequency of species~$1$ be~$f(t,x)$, where~$t$ is time, and~$x$ is position. The frequency of the other species is then~$1-f(t,x)$. In the limit of weak selection, when~$a_{ij}\ll1$, we find a continuum description of this one-dimensional stepping stone model in terms of a generalized stochastic Fisher equation~\cite{korolev:review}

\begin{equation}
\label{eq:dynamics}
\frac{\partial f}{\partial t}=sf(1-f)(f^{*}-f)+D_{s}\frac{\partial^{2}f}{\partial x^{2}}+\sqrt{D_{g}f(1-f)}\Gamma(t,x),
\end{equation} 

\noindent where~$\Gamma(t,x)$ is an It\^{o} delta-correlated Gaussian white noise\footnote{In It\^{o}'s formulation,~$f(t_{0},x)$ and~$\Gamma(t_{0},x)$ are independent for any given~$t_{0}$, but special rules of It\^{o} calculus must be used to differentiate a composite function; see Ref.~\cite{korolev:review} for further discussion.},~$D_{s}$ is the spatial diffusion constant, and~$D_{g}\sim 1/N$ is the strength of number fluctuations. The key parameters~$s=(\alpha_{1}+\alpha_{2})/\tau$ and~$f^{*}=\alpha_{1}/s$ are given by~$\alpha_{1}=(a_{12}-a_{22})/\tau$ and~$\alpha_{2}=(a_{21}-a_{11})/\tau$. The selective advantage (or strength) of mutualism is given by~$s$, while~$f^{*}$ is the equilibrium fraction of species~$1$ that would occur in a spatially homogeneous population without number fluctuations.

The usual mean-field treatment neglects spatial correlations and fluctuations. With the neglect of the last two terms, Eq.~(\ref{eq:dynamics}) becomes an ordinary differential equation, and its dynamics can be easily analyzed. There are four possible outcomes as shown in Fig.~3a and the Appendix. The population develops mutualism when~$\alpha_{1}$~and~$\alpha_{2}>0$, one of the species outcompetes the other when~$\alpha_{1}\alpha_{2}<0$, and, when~$\alpha_{1}$~and~$\alpha_{2}<0$, the population is bistable, with the either species capable of outcompeting the other depending on the initial conditions.

\begin{figure*}
\begin{tabular}{lll}
(a)\includegraphics[height=5cm]{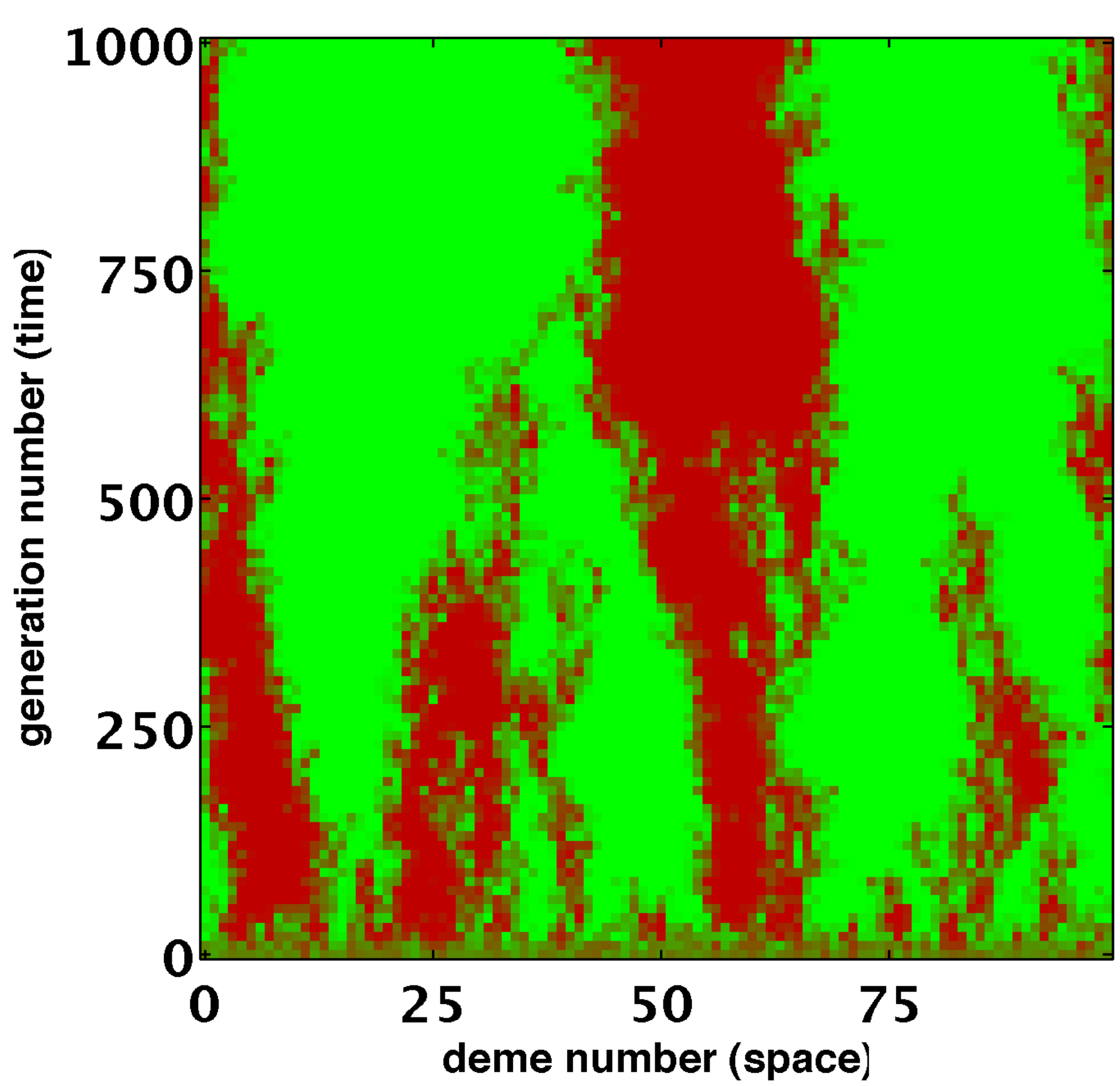} & (b)\includegraphics[height=5cm]{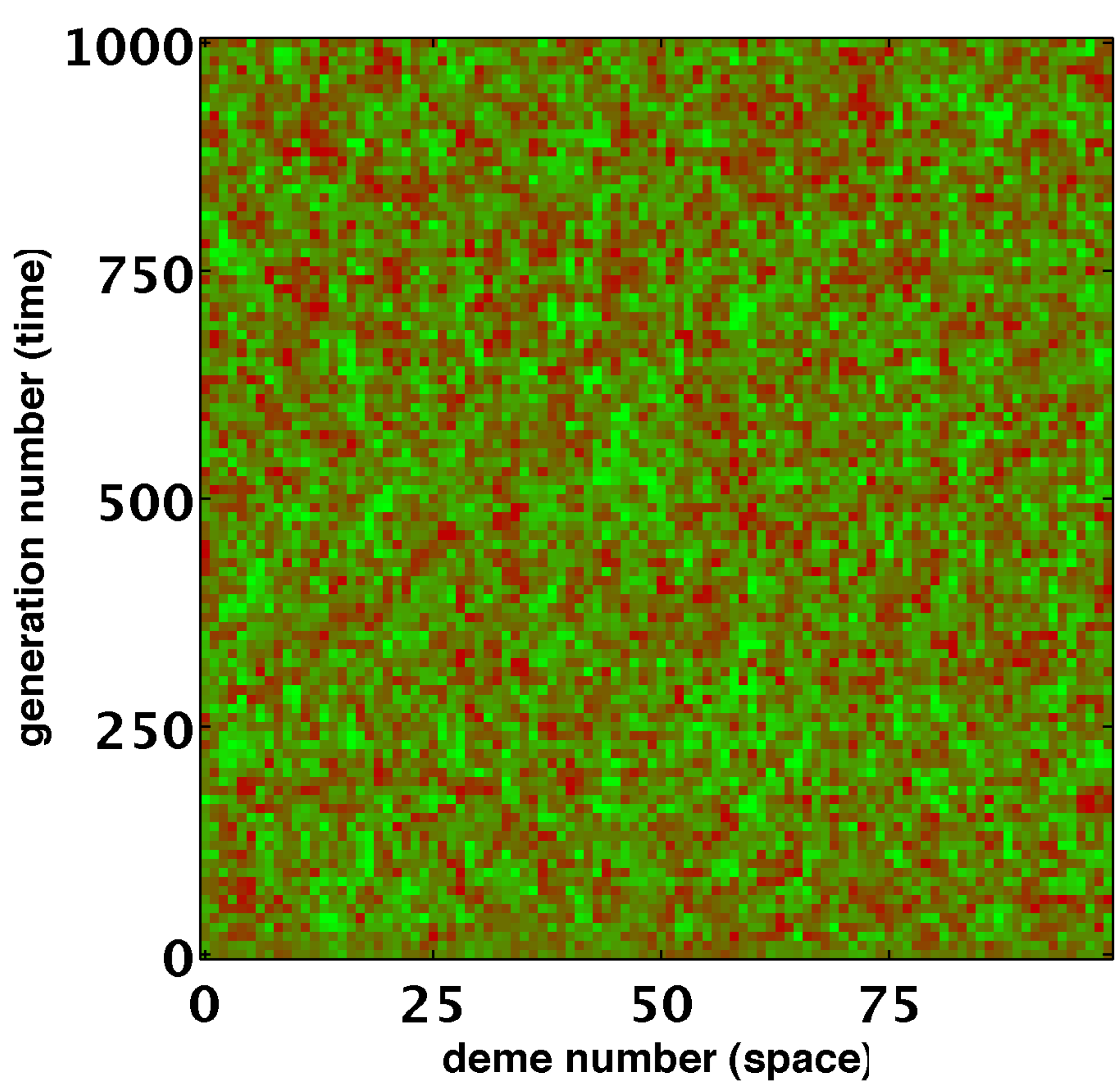} & (c)\includegraphics[height=5cm]{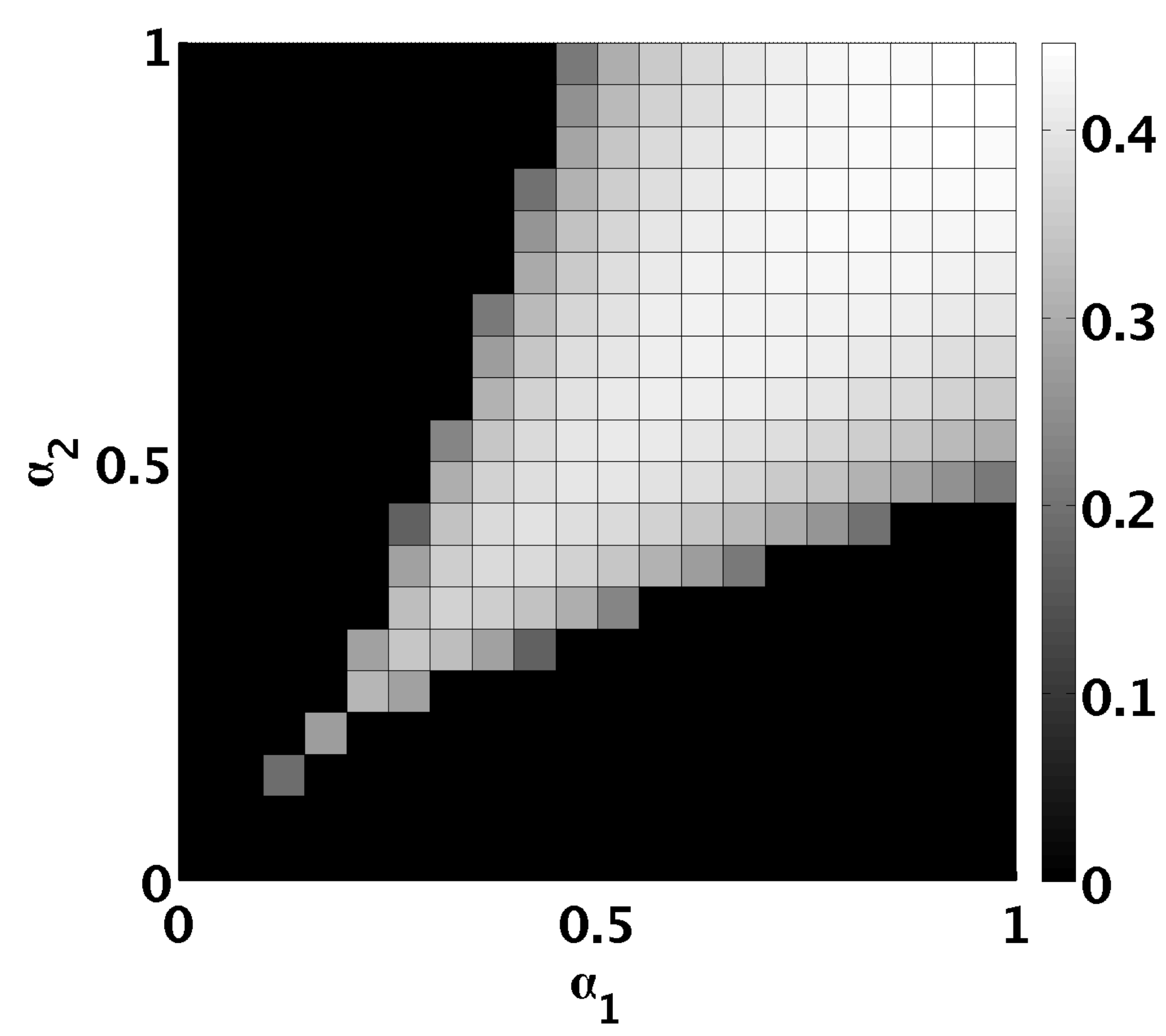}
\end{tabular}
\caption{(Color online) Mutualism in the one-dimensional stepping stone model. (a) Spatial demixing for~$N=30$, $mN=1$, and no interspecies interactions, all $a_{ij}=0$. Green~(light gray) and red~(dark gray) represent species~$1$ and~$2$ respectively. Every deme and every tenth generation are shown. (b) The same as in a, but with strong mutualism~$a_{12}=a_{21}=0.5$. (c) Heat map of~$H(4\cdot10^{6},0)$ from simulations with the same parameters as in a, but with~$10^{4}$ demes and varying~$a_{12}$ and~$a_{21}$.}
    \label{fig:phase_diagrams}
\end{figure*}

We first explore number fluctuations in a population without spatial structure, e.g. consisting of one deme or a finite population of~$N_{\rm{tot}}$ organisms with very large~$D_{s}$. A stochastic treatment must also account for absorbing boundary conditions at~$f=0$ and~$f=1$, when one of the two species goes extinct. The absorbing boundaries arise because there is a finite probability to find the population in any of its discrete states when the population size is finite. Therefore, after a sufficiently long time, the population will reach one of the absorbing boundaries and become fixed. The splitting probabilities and fixation times can be calculated for this zero-dimensional problem with fluctuations using the Kolmogorov backward equations; see the Appendix.

In a spatially extended population, however, local extinctions can be prevented or rescued through migration. Suppose, in particular, that migrations are frequent and mutualism is sufficiently strong to keep the population near the equilibrium fraction~$f^{*}$. In this limit, we can extend the mean-field approximation to account for fluctuations and spatial degrees of freedom by replacing the nonlinear reaction term in Eq.~(\ref{eq:dynamics}) with a linear one:

\begin{equation}
\label{eq:dynamics_linear}
\frac{\partial f}{\partial t}=s f^{*}(1-f^{*})(f^{*}-f)+D_{s}\frac{\partial^{2}f}{\partial x^{2}}+\sqrt{D_{g}f(1-f)}\Gamma(t,x).
\end{equation} 

\noindent If~$f(t,x)\approx f^{*}$, the error we make should be small; more importantly, Eq.~(\ref{eq:dynamics_linear}) can now be solved exactly. The solution is most easily obtained in terms of the average spatial heterozygosity~$H(t,x)$, a two-point correlation function equal to the probability to sample two different species distance~$x$ apart:

\begin{equation}
H(t,x)=\bm{\langle} f(t,0)[1-f(t,x)]+f(t,x)[1-f(t,0)] \bm{\rangle}.
\label{eq:H}
\end{equation}

\noindent Using the It\^{o} calculus, we derive the equation of motion for~$H(t,x)$ from Eq.~(\ref{eq:dynamics_linear}), 

\begin{equation}
\label{eq:H_dynamics}
\frac{\partial H}{\partial t}=\left[2D_{s}\frac{\partial^{2}}{\partial x^{2}}-D_{g}\delta(x)-s H^{*}(1-H^{*})\right]H,
\end{equation}

\noindent where~$\delta(x)$ is the delta function,~$H^{*}=2f^{*}(1-f^{*})$, and we, for simplicity, assumed that the population is initially uniform with the equilibrium fraction~$f(0,x)=f^{*}$. The stationary solution, valid at long times, reads

\begin{equation}
\label{eq:H_stationary_solution}
\frac{H(\infty,x)}{H^{*}}=1-\frac{e^{-x\sqrt{{s f^{*}(1-f^{*})}/{D_{s}}}}}{1+\sqrt{{8s D_{s}f^{*}(1-f^{*})}/{D_{g}^{2}}}}.
\end{equation}

\noindent Since~$H^{*}$ is the heterozygosity of a well-mixed population with~$f=f^{*}$, the fraction on the right hand side is the correction to the mean-field analysis. Thus, we see that, for~$s\ll D_g^{2}/D_{s}$, the probability of the two species coexisting at any particular point in space~[given by~$H(\infty,0)$] becomes small, which is inconsistent with mutualism and our assumption that~$f(t,x)\approx f^{*}$. Hence, we anticipate a critical value of~$s$ below which mutualism \textit{must} give way to spatial demixing.

Although the hierarchy of moment equations does not close for the original, nonlinear problem given by Eq.~(\ref{eq:dynamics}), the average spatial heterozygosity~$H(t,x)$ is still useful for characterizing the behavior of the system. In particular, the average local heterozygosity~$H(t,0)$ can be used to measure the amount of mutualism. Equation~(\ref{eq:H_stationary_solution}) suggests that~$H(t,0)$ reaches a nonzero steady state value when~$s\gg D_{g}^{2}/D_{s}$. However, when~${s}=0$, the exact solution of Eq.~(\ref{eq:dynamics}) reveals that instead of reaching a steady state, $H(t,0)$ decays to zero as~$t^{-1/2}$~\cite{korolev:review}. When species do not coexist locally, mutualism is impossible. Hence, we can use the long time behavior of~$H(t,0)$ to distinguish between populations where mutualism can and cannot persist. See the Appendix for another quantity to distinguish the phases, similar to the susceptibility in equilibrium physics.

The phase digram obtained from simulations is shown in Fig.~\ref{fig:phase_diagrams}c. The region of parameters where mutualism can evolve is significantly reduced compared to the well-mixed prediction shown in Fig~3. In particular, mutualism is impossible even for positive~$s$, provided~$s$ is small. Fluctuations and spatial structure also favor \textit{symmetric} mutualism, with~$\alpha_{1}\approx\alpha_{2}$, i.e. when the two species benefit equally from the interaction. The mutualistic phase~[characterized by~$\lim_{t\to\infty}H(t,0)\neq0$] is separated from the demixed phase~[$\lim_{t\to\infty}H(t,0)=0$] by two lines of second order phase transitions that meet in a cusp: $\lim_{t\to\infty}H(t,0)$ decreases continuously to zero as these lines are approached.

\begin{figure*}
\begin{tabular}{lll}
(a)\includegraphics[height=6cm]{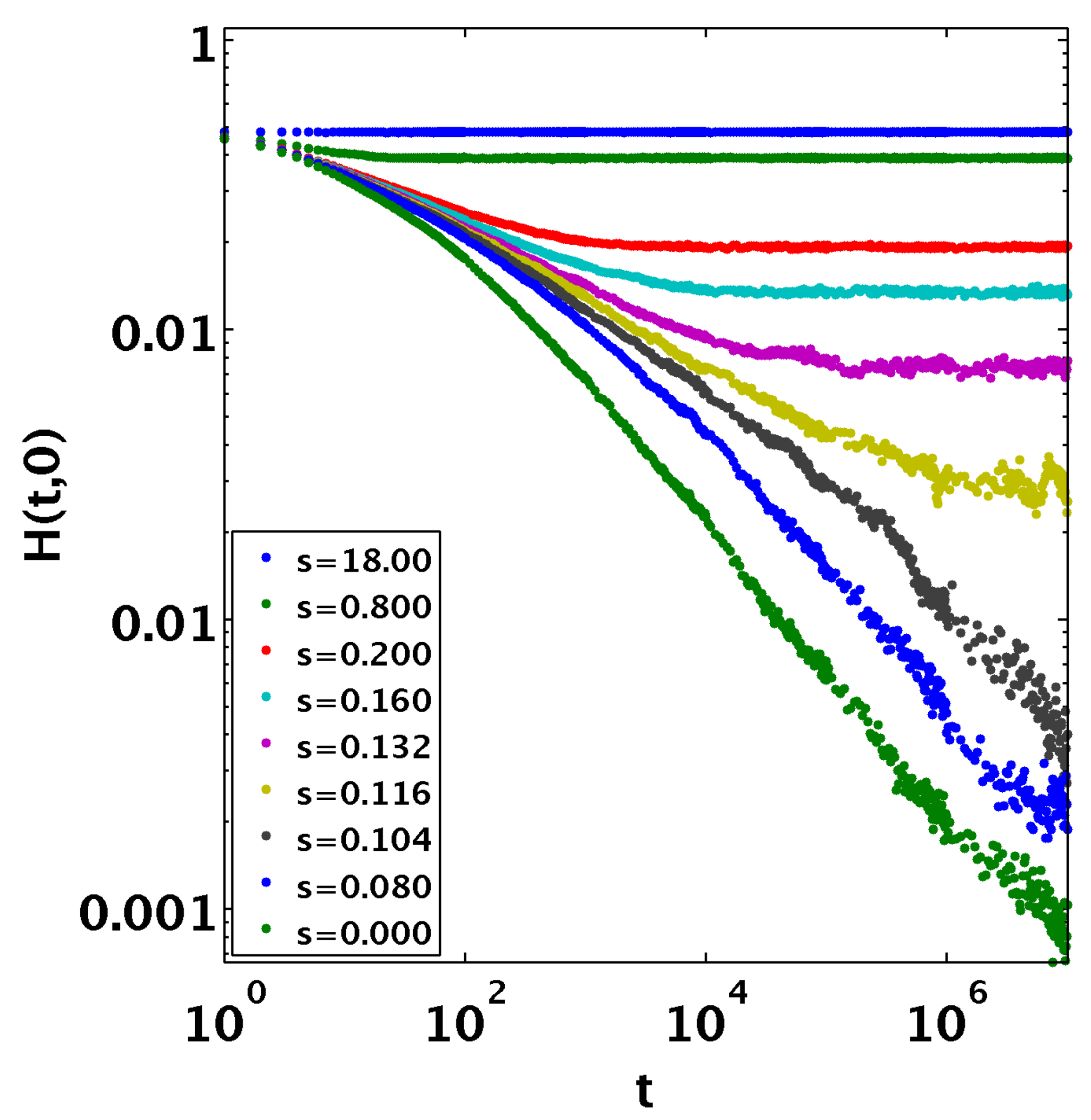} & (b)\includegraphics[height=6cm]{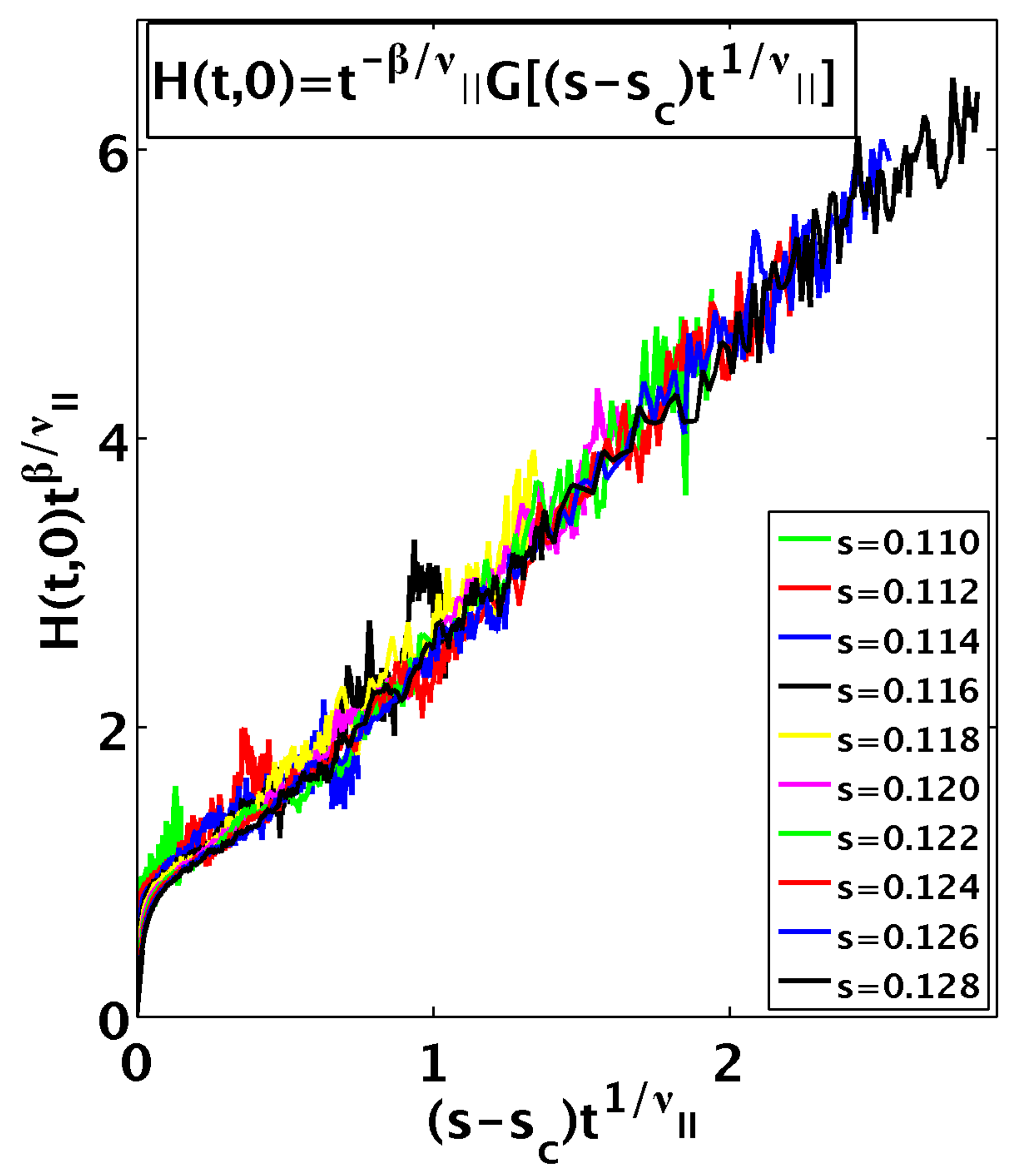} & (c)\includegraphics[height=6cm]{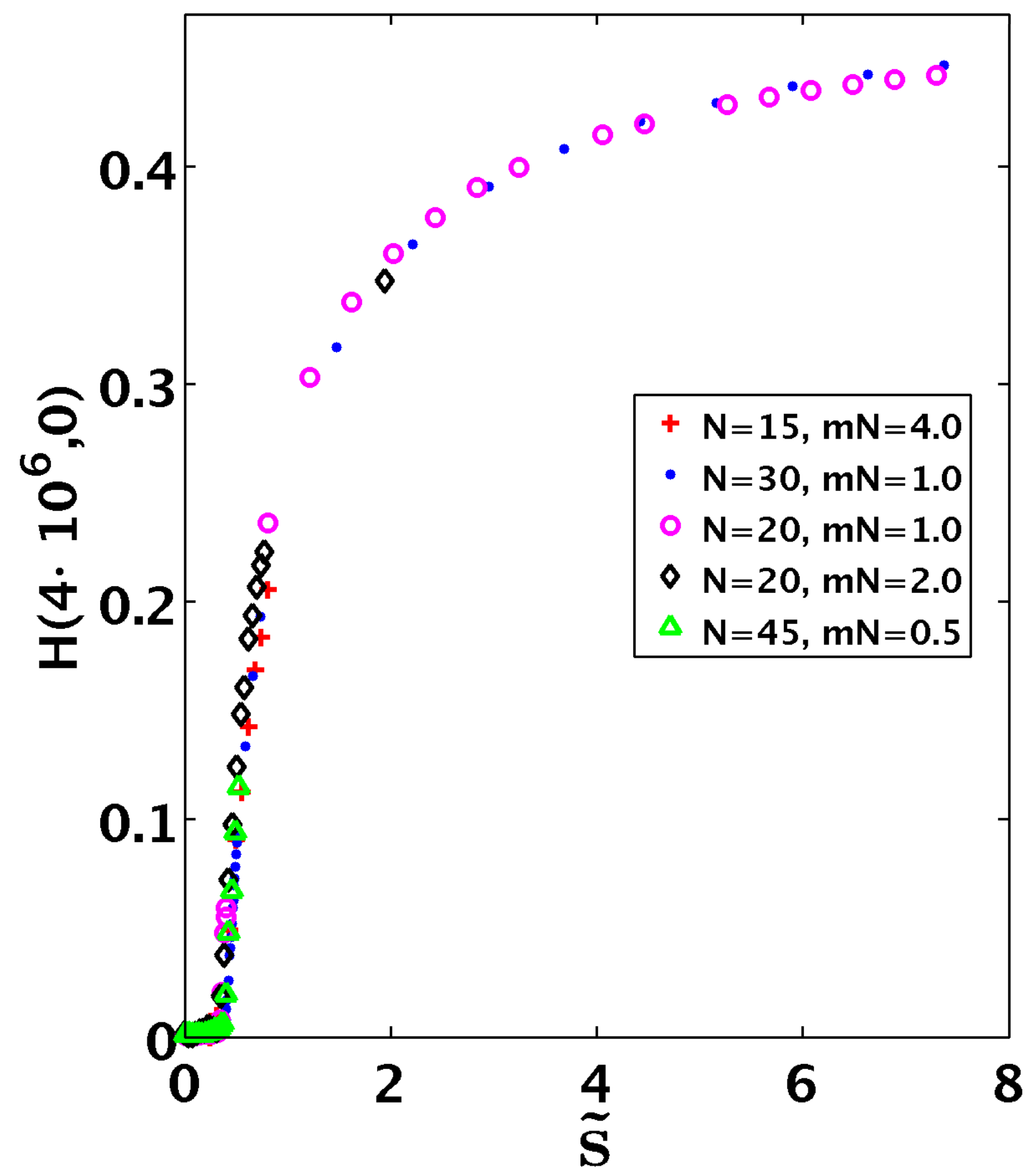}
\end{tabular}
\caption{(Color online) Properties of the DP2 phase transition~($\alpha_{1}=\alpha_{2}$) from simulations with the same parameters as in Fig.~\protect{\ref{fig:phase_diagrams}}c. (a) Decay of the local heterozygosity with time; lower values of the mutualism selective advantage~$s$ lead to faster decay. (b) The collapse of~$H(t,0)$ onto a universal scaling function shown in the inset with DP2 critical exponents for different~$s$ above the phase transition at~$s_{c}=0.109$. For the DP2 critical point, we expect~$\beta=0.92$ and~$\nu_{\parallel}=3.22$~\protect{\cite{hinrichsen:review}}.  (c) $H(t=4\cdot10^{6},0)$ as a function of~$\tilde{s}=sD_{s}/D_{g}^{2}$ for different values of~$m$ and~$N$. We show how to obtain~$D_{s}/D_{g}^{2}$ in the Appendix.}
    \label{fig:plots}
\end{figure*}

Nonequilibrium phase transitions from an active~(mixed) to an absorbing~(one of the two species) state have been studies extensively; see Ref.~\cite{hinrichsen:review}. Generically, when the absorbing states are not symmetric,~$\alpha_{1}\neq\alpha_{2}$, the exit from mutualism belongs to the DP universality class. We can most readily see this for~$f^{*}$ close to an absorbing boundary, say~$f^{*}\ll1$. For large mutualistic selective advantage~$s$, species~$1$ then remains at low frequencies throughout the population. As the strength of mutualism is decreased, some spatial regions stochastically lose species~$1$, but the more abundant species~$2$ persists. Local extinctions are opposed by the spread of species~$1$ from the nearby regions via Fisher waves. This dynamics is just that of DP in~\cite{hinrichsen:review}. When~$\alpha_{1}=\alpha_{2}$, the absorbing states are symmetric and the local extinctions of either species are equally likely. As a result, this phase transition belongs to DP2 universality class. We checked that our simulations are consistent with the DP2 ``bicritical point'' by calculating how~$H(t,0)$ decays for different values of~$s$ in a population that is initially well-mixed (see Fig.~\ref{fig:plots}a) and then collapsing these decay curves onto a unique scaling function using DP2 exponents as shown in Fig.~\ref{fig:plots}b. Equation~(\ref{eq:dynamics}) is also known to describe a DP2 transition for~$f^{*}=1/2$~\cite{hammal:langevin}. Although the DP2 transition occurs only at a point, it influences a large portion of the phase diagram and governs the nonlinear shape of the DP transition lines near this ``bicritical point.''

To understand how phase boundaries depend on the parameters of the model, it is convenient to measure distance in the units of~$D_{s}/D_{g}$ and time in the units of~$D_{s}/D_{g}^{2}$. When, Eq.~(\ref{eq:dynamics}) is nondimensionalized, and the dynamics is controlled by only two dimensionless parameters,~$f^{*}$ and~$\tilde{s}=s D_{s}/D_{g}^{2}$. We confirm this data collapse in simulations, see Fig.~\ref{fig:plots}c.

Spatial structure and number fluctuations change not only the mutualistic region~($\alpha_{1},\alpha_{2}>0$), but also the whole phase diagram for well-mixed populations, see Fig~3. In particular, there is no bistable phase~(with splitting probabilities sensitive to the initial conditions) in 1d spatial populations. For almost all initial conditions, domains of species~$1$ or species~$2$ appear because of number fluctuations; the subsequent behavior can be analyzed in terms of the Fisher wave velocities of the domain boundaries. For~$\alpha_{1}>\alpha_{2}$, this velocity is directed from species~$1$ to species~$2$, and the direction is reversed for~$\alpha_{1}<\alpha_{2}$. As a result, one of the species takes over, much like an equilibrium first order phase transition proceeds through nucleation and growth.

When~$\alpha_{1}=\alpha_{2}$ and mutualism is unstable, population segregates into single species domains, and the dynamics is driven by the random walks of domain boundaries. For~$s>0$ this demixing is slowed down by mutualism, but for~$s<0$ it is \textit{sped up} initially due to the reaction term in Eq.~(\ref{eq:dynamics}). After domains form, however, more negative values of~$s$ lead to slower coarsening because the diffusion constant of domain boundaries decreases. Surprisingly, the exactly solvable limit of~$\alpha_{1}=\alpha_{2}=0$ undergoes the fastest demixing in the long time limit for large system sizes (see Fig.~4).

We have shown that a critical strength of mutualism is required to overcome the spatial demixing of species driven by local number fluctuations. The critical strength of mutualism strongly depends on the symmetry of the interaction, and mutualism is more likely to evolve between species that share the benefits equally. We believe that these predictions can be tested in growing microbial colonies on a Petri dish because colony frontiers behave as quasi-one-dimensional populations~\cite{korolev:review} and because mutualistic interactions can be engineered in the lab~\cite{gore:sucrose}. Such experimental studies would be of value not only for evolutionary biology, but also for nonequilibrium statistical mechanics because they could provide important experimental tests of the theory.

After submitting this paper, we learned of a preprint by Dall'Astra~\textit{et al.}, ``Strong noise effects in one-dimensional neutral populations'' (arXiv:1012.1209), which discusses symmetric cooperation in a similar model, corresponding to~$f^{*}=1/2$ and~$\alpha_{1}=\alpha_{2}$ in our terminology. Our work was constructed with experiments at microbial frontiers in mind;  hence, it differs due to its focus on asymmetric interactions~$\alpha_{1}$ and~$\alpha_{2}$ of arbitrary sign and the large deme sizes in our simulations.

\begin{acknowledgments}
We are indebted to A.~Murray and M.~Mueller for interesting us in mutualism and for frequent discussions about the experimental situation. Work supported in part by the National Science Foundation through Grant No. DMR-1005289 and by the Harvard Materials Research Science and Engineering Center through NSF Grant No. DMR-0820484. KSK is also supported by a Pappalardo fellowship at MIT.
\end{acknowledgments}

\appendix

\section{Supplementary Material}

\begin{figure*}[h!]
\begin{tabular}{ll}

(a)\includegraphics[height=0.7\columnwidth]{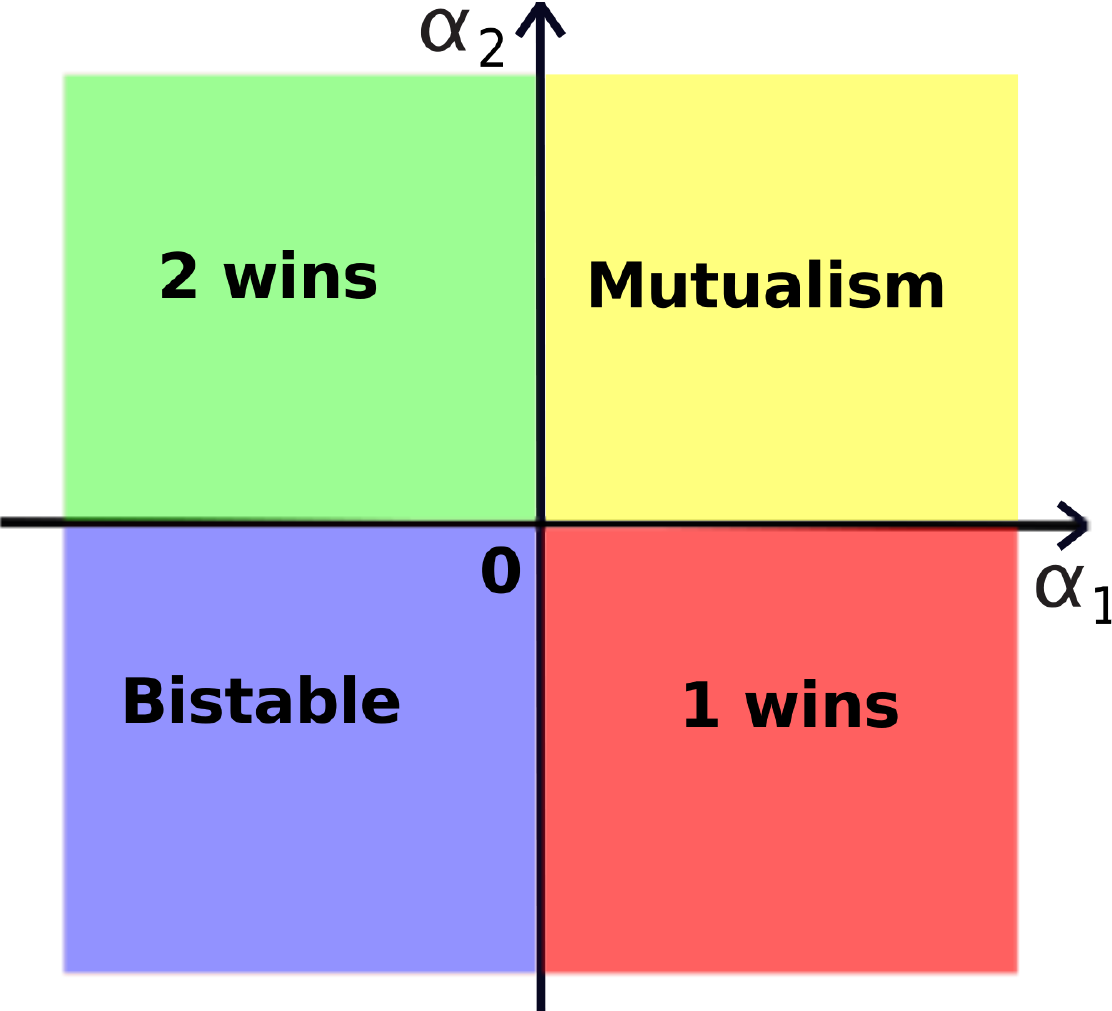} & (b)\includegraphics[height=0.7\columnwidth]{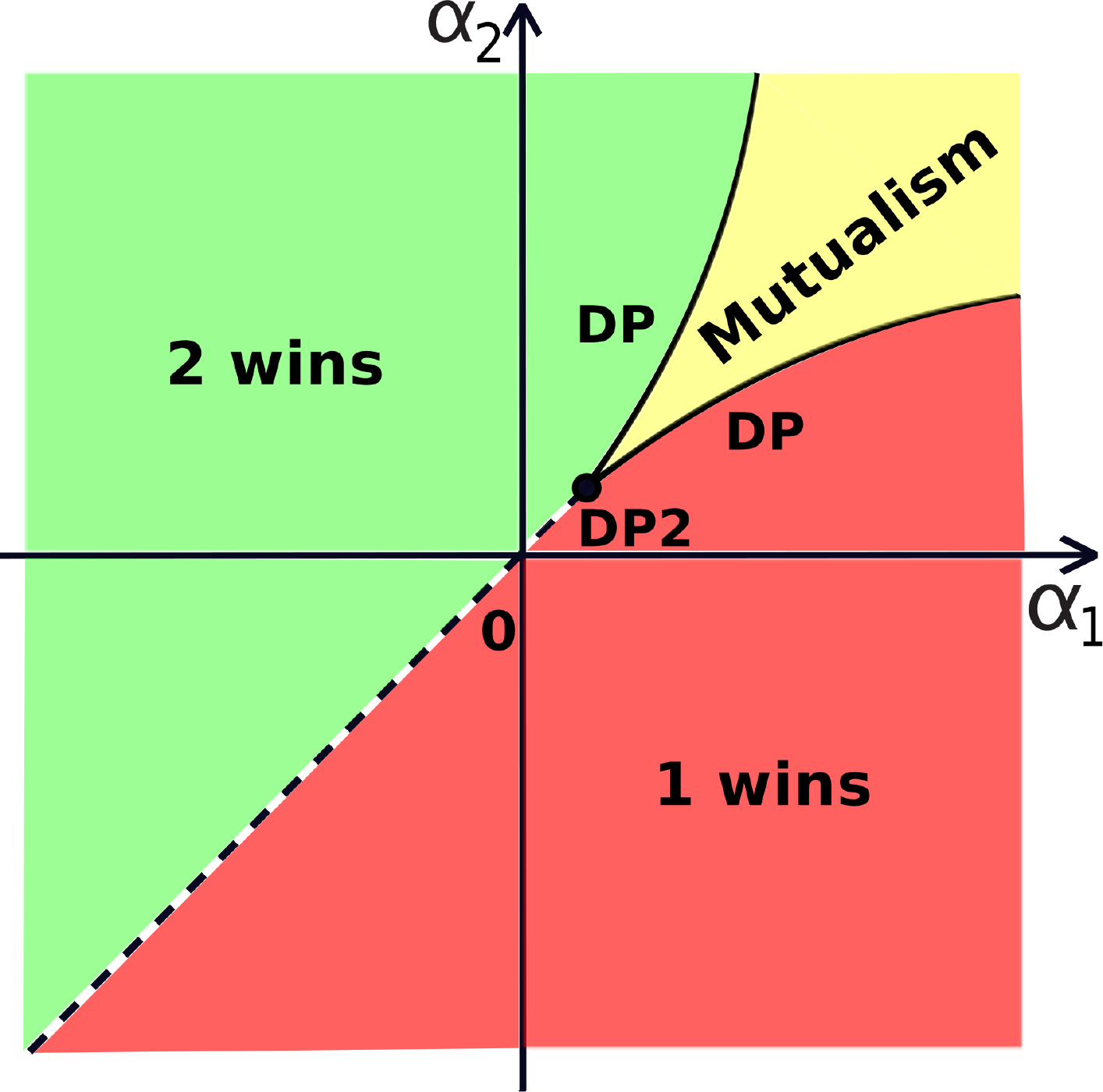} \\

(c)\includegraphics[width=0.7\columnwidth]{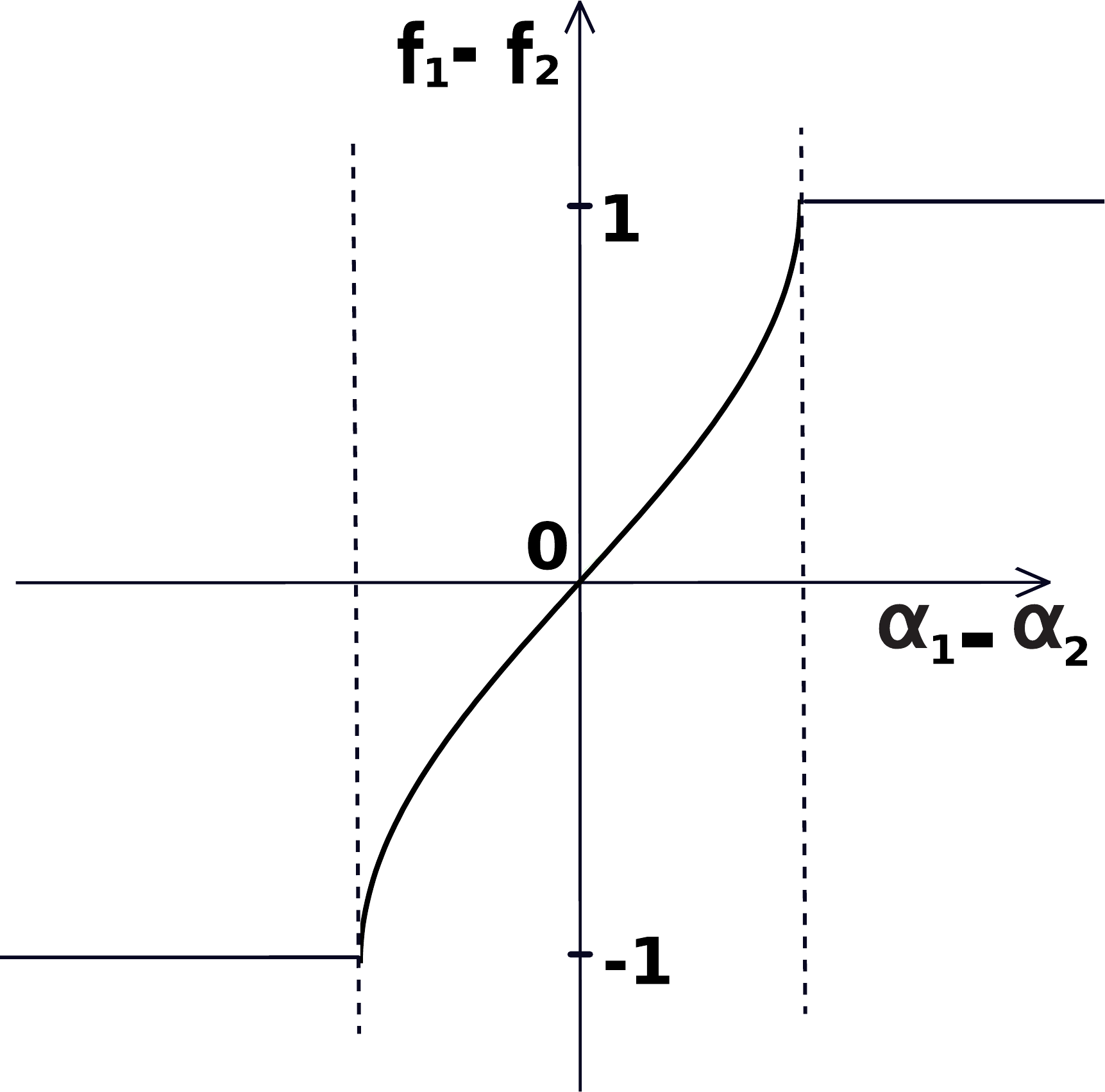} & (d)\includegraphics[width=0.7\columnwidth]{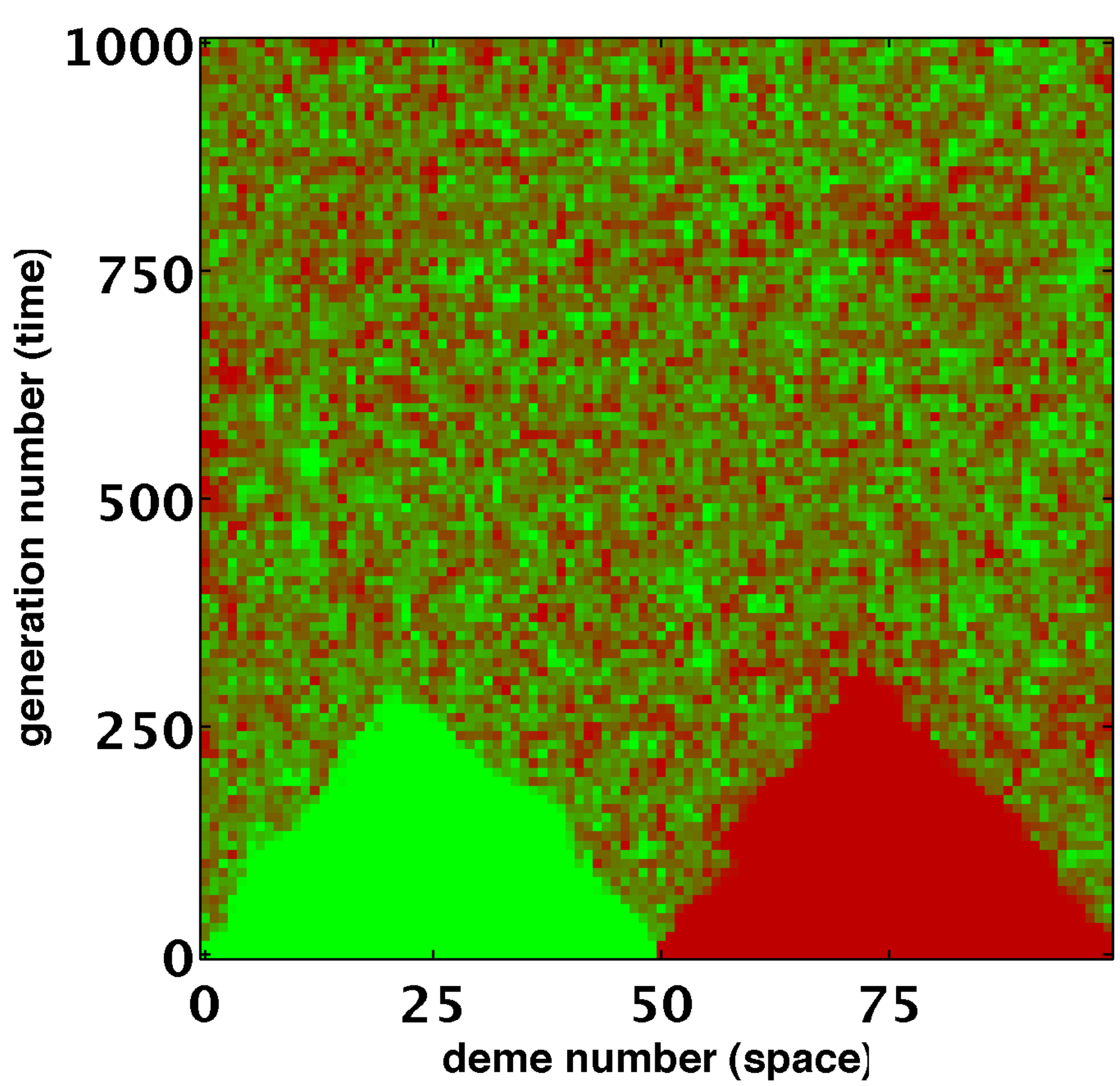} \\

\end{tabular}
\caption{(Color online) Phase diagrams of cooperation and competition. (a) The phase diagram in the mean-field limit, which neglects spatial structure and number fluctuations. (b) Schematic phase diagram for evolutionary games in one dimension. (c) Schematic plot of the change of species equilibrium frequencies along a cut~$\alpha_{1}+\alpha_{2}=\rm{const}$ intersecting the DP phase transition lines through the yellow region in the phase diagram shown in (b). The two dashed lines show the locations of the DP phase transitions. We find that the nonlinear dependence of the~$f_{1}-f_{2}$ on~$\alpha_{1}-\alpha_{2}$ near DP transitions is consistent with the DP exponent~$\beta=0.276486$~\protect{\cite{hinrichsen:review}}. (d) Illustration of the fact that the phase behavior summarized in (a)-(c) is not specific to the well-mixed initial conditions used for most of our simulations. Green~(light gray) and red~(dark gray) represent species~$1$ and~$2$ respectively. The simulation parameters correspond to the yellow, mutualistic phase in (b). In fact, the same parameters are used as in Fig.~1b, but with the different initial conditions. At~$t=0$, the population consists of two equal domains: one of species~$1$ and one of species~$2$. Due to the periodic boundary conditions, this initial condition leads to two domain boundaries. The system rapidly forms a mutualistic phase, which squeezes out the red and green domains.}
\end{figure*}


\subsection*{Competition and cooperation in a zero-dimensional model}
Here, we present a more detailed analysis of evolutionary game theory in a spatially homogeneous~(well-mixed) population with~$N_{\rm{tot}}$ organisms. We consider only two species, so~$f(t)$, the frequency of species~$1$, is sufficient to fully characterize the state of the population. The equation of motion for~$f(t)$ takes the following form

\begin{equation}
\frac{d}{dt}f(t)=s f(t)[1-f(t)][f^{*}-f(t)]+\sqrt{\mathcal{D}_{g}f(t)[1-f(t)]}\Gamma(t),
\label{eq:wmdynamics}
\end{equation}

\noindent where~$\Gamma(t)$ is a zero mean Gaussian white noise interpreted according to the It\^{o} prescription.

\begin{equation}
\bm{\langle} \Gamma(t_{1})\Gamma(t_{2}) \bm{\rangle}=\delta(t_{1}-t_{2}),
\end{equation}

\noindent and~$\mathcal{D}_{g}\sim2/(N_{\rm{tot}}\tau)$ measures the strength of number fluctuations.

In the absence of noise, the system has four generic types of behavior.

\begin{itemize}

\item $\alpha_{1}>0$, $\alpha_{2}>0$ corresponds to the snowdrift game or mutualism~\cite{nowak:book}. The stable fixed point is at~$f=f^{*}$, which is between zero and one.

\item $\alpha_{1}<0$, $\alpha_{2}<0$ corresponds to the coordination game. There are two stable fixed points~$f=0$ and~$f=1$. $f^{*}$ is still between zero and one, but this fixed point is unstable. The long time fate of the system is entirely determined by whether the initial fraction of species~$1$ is above or below~$f^{*}$.

\item $\alpha_{1}>0$, $\alpha_{2}<0$ corresponds to competitive exclusion. Species~$1$ dominates, and the stable fixed point is at~$f=1$. 

\item $\alpha_{1}<0$, $\alpha_{2}>0$ also corresponds to competitive exclusion. Species~$2$ dominates, and the stable fixed point is at~$f=0$. 

\end{itemize}

In the presence of noise, the system always reaches one of the absorbing states at~$f=0$ or~$f=1$ at long times. The splitting probabilities can be calculated from the backward Kolmogorov equation~\cite{risken:FPE} , and we find that~$U(f_{0})$, the probability to eventually reach~$f=1$ starting from~$f=f_{0}$, is given by

\begin{equation}
\label{eq:splitting_probabilities}
U(f_{0})=\frac{\int_{0}^{f_{0}}e^{\frac{s}{\mathcal{D}_{g}}(f-f^{*})^{2}}df}{\int_{0}^{1}e^{\frac{s}{\mathcal{D}_{g}}(f-f^{*})^{2}}df}.
\end{equation} 

\noindent In the limit of strong noise~$\mathcal{D}_{g}\gg s$, the fixation probability is approximately given by

\begin{equation}
U(f_{0})=f_{0}[1+\frac{s}{3\mathcal{D}_{g}}(1-3f^{*}+3f_{0}-f_{0}^{2})].
\end{equation}

\noindent Note that for~$f\ll f^{*}$ we recover the~$1/3$-law~\cite{nowak:finite}, which states that, for~$s<0$ the fixation probability of species~$1$ is greater than it would be under purely stochastic~(neutral) dynamics, provided~$f^{*}>1/3$.  

The average time to reach fixation can be calculated; see~\cite{risken:FPE}. Cremer, Reichenbach, and Frey~\cite{cremer:stochastic} have recently carried out the analysis of fixation time and found that it scales as~$1/\mathcal{D}_{g}$ for~$|s|\ll \mathcal{D}_{g}$, but becomes exponentially large when mutualism is strong.

Another way to analyze the effects of noise on the dynamics is to look at the Fokker-Plank equation corresponding to~Eq.~(\ref{eq:wmdynamics}), namely 

\begin{equation}
\begin{aligned}
\frac{\partial}{\partial t}P(t,f)=&-s\frac{\partial}{\partial f}\left[f(1-f)(f^{*}-f)P(t,f)\right]+\\ &\frac{D_{g}}{2}\frac{\partial^{2}}{\partial f^{2}}\left[f(1-f)P(t,f)\right]
\end{aligned}
\end{equation}

\noindent where~$P(t,f)$ is the probability distribution of~$f$ at time~$t$. The Fokker-Plank equation is more intuitive for most physicists when the diffusion coefficient does not depend on~$f$ because, in this case, it describes diffusion of a particle in a potential. We can achieve such a simplification by the following change of variables

\begin{equation}
\label{eq:change_variables}
f=\sin^{2}(p/2).
\end{equation}

\noindent The potential is then given by

\begin{equation}
\label{eq:potential}
V=\mathcal{D}_{g}\ln[\sin(p)]+\frac{s}{2}\cos(p)\left(f^{*}-\frac{1}{2}\right)+\frac{s}{8}\cos^{2}(p).
\end{equation}

\noindent For~$f^{*}=1/2$, $V$ has a minimum in the interior only for~$s>4\mathcal{D}_{g}$ suggesting that the decay of weak mutualism is not protected by a barrier, which is also true for other values of~$f^{*}$.

\subsection*{Connection between~$\bm{H(t,x)}$ and a susceptibility}

In the main text, we use~$\lim_{t\to\infty}H(t,0)$ to distinguish between mutualistic and non-mutualistic phases. Although this quantity is very convenient and easy to measure in simulations, it is rarely used in equilibrium physics to characterize phase transitions. We can make a more explicit connection between the equilibrium theory of phase transitions and the out of equilibrium loss of mutualism by considering the following quantity

\begin{equation}
\label{eq:chi}
\chi(t)=-\int_{-\infty}^{\infty}[H(t,x)-\lim_{x\to\infty}H(t,x)]dx.
\end{equation} 

\noindent Note that~$H(t,x)<\lim_{x\to\infty}H(t,x)$, so~$\chi$ is positive and measures the degree of species demixing. Since $\chi$ is an integral of a two-point correlation function~$H(t,x)$, it is analogous to the susceptibility used in equilibrium physics. 

In the mutualistic phase, both species are present, so~$\lim_{t\to\infty}\lim_{x\to\infty}H(t,x)>0$. Then~$\lim_{t\to\infty}\chi(t)$ is finite and measures the degree and spatial extent of the reduction of the species diversity due to number fluctuations. Close to one of the phase transition lines in Fig.~3b,~$\lim_{t\to\infty}H(t,0)\approx0$; therefore,~$\lim_{t\to\infty}\chi(t)\approx\xi\lim_{x\to\infty}H(t,x)$, where~$\xi$ is the correlation length. As the phase transition is approached,~$\xi$ diverges with an exponent~$\nu_{\perp}$, which is~$1.096854$ and~$1.83$ for DP and DP2 transition respectively. In addition,~$\lim_{x\to\infty}H(t,x)=1/2$ for the DP2 transition, but~$\lim_{x\to\infty}H(t,x)$ vanishes with exponent~$\beta=0.276486$ for the DP transition~\cite{hinrichsen:review}.

In a non-mutualistic phase,~$\chi(t)$ diverges with time because~$H(t,x)\approx0$ for~$x$ smaller than the typical domain size, which grows in time due to coarsening. Close to the phase transition this divergence occurs with the exponent~$1/z$, where the dynamical exponent~$z=1.580745$ and~$z=1.74$ for DP and DP2 transitions respectively~\cite{hinrichsen:review}. For the exactly solvable case of~$s=0$,~$\chi(t)\sim t^{1/2}$.

\subsection*{Estimating $D_{g}$ and $D_{s}$}

Generalized stochastic Fisher-Kolmogorov equations, e.g. Eq.~(1), describe population dynamics on long time and length scales; therefore, the diffusion constant~$D_{s}$ and the strength of number fluctuations~$D_{g}$ should be viewed as phenomenological parameters in an effective field theory. As a result, they may often be complicated functions of the microscopic parameters. For the stepping stone model,~$D_{s}=ma^{2}/(2\tau)$ and~$D_{g}=2a/(N\tau)$ in the limit of large~$N$ and small~$m$~\cite{korolev:review}. Our computer simulations, however, are not in this limit, and we have to estimate~$D_{s}$ and~$D_{g}$ numerically. 

This estimation can be easily done from simulations with all~$a_{ij}=0$ because the stepping stone model without species interactions can be solved exactly. In particular, the decay of~$H(t,0)$ from a homogeneous 1:1 mixture of the species satisfies the following relation~\cite{korolev:review}    

\begin{equation}
\label{EH0LimitNoMutation}
H(t,0)=\left(\frac{\pi D_{g}^2 t}{2D_{s}}\right)^{-1/2}+O(t^{-3/2}).
\end{equation}

\noindent We can then obtain~$D_{s}/D_{g}^{2}$ by fitting Eq.~(\ref{EH0LimitNoMutation}) to the simulation results. One can also estimate~$D_{s}$ and~$D_{g}$ individually from the average size of domains; see Ref.~\cite{korolev:review}.

In our letter, we used the aforementioned estimate of~$D_{s}/D_{g}^{2}$ to collapse~$H(t=\infty,x=0)$ as a function of~$\tilde{s}$ in Fig.~2c. We found that the DP2 transition occurs at~$\tilde{s}=0.80$.


\begin{figure}[h!]
\caption{(Color online) Decay of $H(t,0)$ in initially well-mixed populations with~$N=30$, $mN=1$, $a_{11}=a_{22}=0$, $a_{12}=a_{21}$, and $10^{4}$ demes. Note that for more negative~$a_{12}$ the initial decay is faster, but it crosses over to a slower decay at longer times.}
\includegraphics[height=\columnwidth]{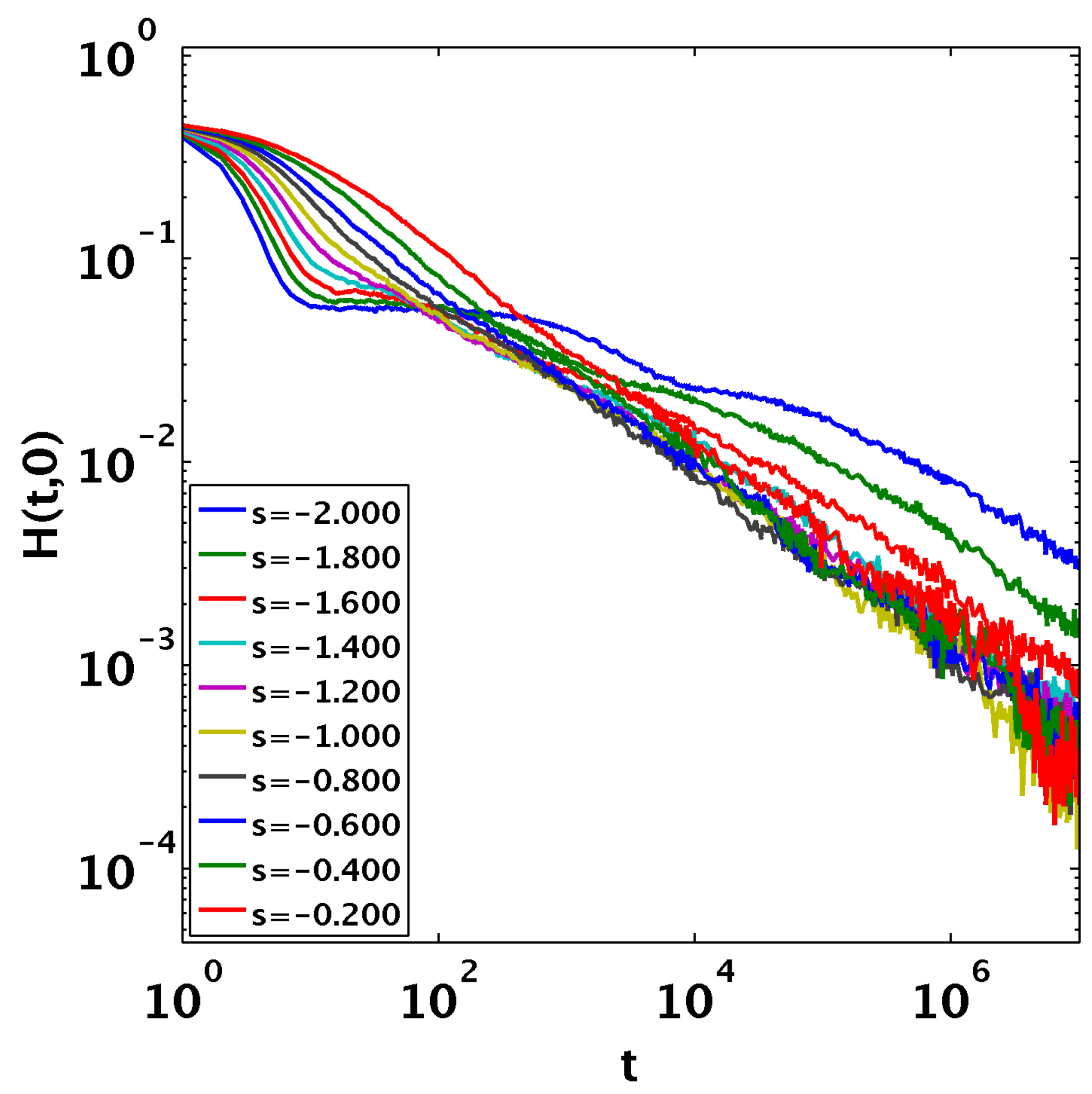}
\end{figure}


\begin{thebibliography}{15}
\expandafter\ifx\csname natexlab\endcsname\relax\def\natexlab#1{#1}\fi
\expandafter\ifx\csname bibnamefont\endcsname\relax
  \def\bibnamefont#1{#1}\fi
\expandafter\ifx\csname bibfnamefont\endcsname\relax
  \def\bibfnamefont#1{#1}\fi
\expandafter\ifx\csname citenamefont\endcsname\relax
  \def\citenamefont#1{#1}\fi
\expandafter\ifx\csname url\endcsname\relax
  \def\url#1{\texttt{#1}}\fi
\expandafter\ifx\csname urlprefix\endcsname\relax\def\urlprefix{URL }\fi
\providecommand{\bibinfo}[2]{#2}
\providecommand{\eprint}[2][]{\url{#2}}

\bibitem[{\citenamefont{Nowak}(2006)}]{nowak:book}
\bibinfo{author}{\bibfnamefont{M.}~\bibnamefont{Nowak}},
  \emph{\bibinfo{title}{{Evolutionary dynamics: exploring the equations of
  life}}} (\bibinfo{publisher}{Belknap Press}, \bibinfo{year}{2006}).

\bibitem[{\citenamefont{{Maynard Smith}}(1982)}]{smith:games}
\bibinfo{author}{\bibfnamefont{J.}~\bibnamefont{{Maynard Smith}}},
  \emph{\bibinfo{title}{{Evolution and the Theory of Games}}}
  (\bibinfo{publisher}{Cambridge University Press, Cambridge UK},
  \bibinfo{year}{1982}).

\bibitem[{\citenamefont{{O. Hallatschek, \textit{et
  al.}}}(2007)}]{HallatschekNelson:ExperimentalSegregation}
\bibinfo{author}{\bibnamefont{{O. Hallatschek, \textit{et al.}}}},
  \bibinfo{journal}{Proc. Natl. Acad. Sci. USA} \textbf{\bibinfo{volume}{104}},
  \bibinfo{pages}{19926} (\bibinfo{year}{2007}).

\bibitem[{\citenamefont{{K. S. Korolev, \textit{et
  al.}}}(2010)}]{korolev:review}
\bibinfo{author}{\bibnamefont{{K. S. Korolev, \textit{et al.}}}},
  \bibinfo{journal}{Rev. Mod. Phys.} \textbf{\bibinfo{volume}{82}},
  \bibinfo{pages}{1691} (\bibinfo{year}{2010}).

\bibitem[{\citenamefont{Nowak and May}(1992)}]{nowak:spatialPD92}
\bibinfo{author}{\bibfnamefont{M.~A.} \bibnamefont{Nowak}} \bibnamefont{and}
  \bibinfo{author}{\bibfnamefont{R.~M.} \bibnamefont{May}},
  \bibinfo{journal}{Nature} \textbf{\bibinfo{volume}{359}},
  \bibinfo{pages}{826} (\bibinfo{year}{1992}).

\bibitem[{\citenamefont{Hauert and Doebell}(2004)}]{hauert:snowdrift}
\bibinfo{author}{\bibfnamefont{C.}~\bibnamefont{Hauert}} \bibnamefont{and}
  \bibinfo{author}{\bibfnamefont{M.}~\bibnamefont{Doebell}},
  \bibinfo{journal}{Nature} \textbf{\bibinfo{volume}{428}},
  \bibinfo{pages}{643} (\bibinfo{year}{2004}).

\bibitem[{\citenamefont{Roca et~al.}(2009)\citenamefont{Roca, Cuesta, and
  Sanchez}}]{roca:spatial_cooperation}
\bibinfo{author}{\bibfnamefont{C.}~\bibnamefont{Roca}},
  \bibinfo{author}{\bibfnamefont{J.}~\bibnamefont{Cuesta}}, \bibnamefont{and}
  \bibinfo{author}{\bibfnamefont{A.}~\bibnamefont{Sanchez}},
  \bibinfo{journal}{Rhys. Rev. E} \textbf{\bibinfo{volume}{80}},
  \bibinfo{pages}{046106} (\bibinfo{year}{2009}).

\bibitem[{\citenamefont{Scheucher and
  Spohn}(1988)}]{Scheucher:SpinodalDecomposition}
\bibinfo{author}{\bibfnamefont{M.}~\bibnamefont{Scheucher}} \bibnamefont{and}
  \bibinfo{author}{\bibfnamefont{H.}~\bibnamefont{Spohn}}, \bibinfo{journal}{J.
  Stat. Phys.} \textbf{\bibinfo{volume}{53}}, \bibinfo{pages}{279}
  (\bibinfo{year}{1988}).

\bibitem[{\citenamefont{Kimura and Weiss}(1964)}]{KimuraWeiss:SSM}
\bibinfo{author}{\bibfnamefont{M.}~\bibnamefont{Kimura}} \bibnamefont{and}
  \bibinfo{author}{\bibfnamefont{G.~H.} \bibnamefont{Weiss}},
  \bibinfo{journal}{Genetics} \textbf{\bibinfo{volume}{49}},
  \bibinfo{pages}{561} (\bibinfo{year}{1964}).

\bibitem[{\citenamefont{Hinrichsen}(2000)}]{hinrichsen:review}
\bibinfo{author}{\bibfnamefont{H.}~\bibnamefont{Hinrichsen}},
  \bibinfo{journal}{Advances in Physics} \textbf{\bibinfo{volume}{49}},
  \bibinfo{pages}{815} (\bibinfo{year}{2000}).

\bibitem[{\citenamefont{Al~Hammal et~al.}(2005)\citenamefont{Al~Hammal,
  Chat{\'e}, Dornic, and Mu{\~n}oz}}]{hammal:langevin}
\bibinfo{author}{\bibfnamefont{O.}~\bibnamefont{Al~Hammal}},
  \bibinfo{author}{\bibfnamefont{H.}~\bibnamefont{Chat{\'e}}},
  \bibinfo{author}{\bibfnamefont{I.}~\bibnamefont{Dornic}}, \bibnamefont{and}
  \bibinfo{author}{\bibfnamefont{M.~A.} \bibnamefont{Mu{\~n}oz}},
  \bibinfo{journal}{Phys. Rev. Lett.} \textbf{\bibinfo{volume}{94}},
  \bibinfo{pages}{230601} (\bibinfo{year}{2005}).

\bibitem[{\citenamefont{Gore et~al.}(2009)\citenamefont{Gore, Youk, and
  Van~Oudenaarden}}]{gore:sucrose}
\bibinfo{author}{\bibfnamefont{J.}~\bibnamefont{Gore}},
  \bibinfo{author}{\bibfnamefont{H.}~\bibnamefont{Youk}}, \bibnamefont{and}
  \bibinfo{author}{\bibfnamefont{A.}~\bibnamefont{Van~Oudenaarden}},
  \bibinfo{journal}{Nature} \textbf{\bibinfo{volume}{459}},
  \bibinfo{pages}{253} (\bibinfo{year}{2009}).

\bibitem[{\citenamefont{Risken}(1989)}]{risken:FPE}
\bibinfo{author}{\bibfnamefont{H.}~\bibnamefont{Risken}},
  \emph{\bibinfo{title}{{The Fokker-Planck equation: Methods of Solution and
  Applications}}} (\bibinfo{publisher}{Springer, Berlin and Heidelberg},
  \bibinfo{year}{1989}).

\bibitem[{\citenamefont{Nowak et~al.}(2004)\citenamefont{Nowak, Sasaki, Taylor,
  and Fudenberg}}]{nowak:finite}
\bibinfo{author}{\bibfnamefont{M.}~\bibnamefont{Nowak}},
  \bibinfo{author}{\bibfnamefont{A.}~\bibnamefont{Sasaki}},
  \bibinfo{author}{\bibfnamefont{C.}~\bibnamefont{Taylor}}, \bibnamefont{and}
  \bibinfo{author}{\bibfnamefont{D.}~\bibnamefont{Fudenberg}},
  \bibinfo{journal}{Nature} \textbf{\bibinfo{volume}{248}},
  \bibinfo{pages}{646} (\bibinfo{year}{2004}).

\bibitem[{\citenamefont{Cremer et~al.}(2009)\citenamefont{Cremer, Reichenbach,
  and Frey}}]{cremer:stochastic}
\bibinfo{author}{\bibfnamefont{J.}~\bibnamefont{Cremer}},
  \bibinfo{author}{\bibfnamefont{T.}~\bibnamefont{Reichenbach}},
  \bibnamefont{and} \bibinfo{author}{\bibfnamefont{E.}~\bibnamefont{Frey}},
  \bibinfo{journal}{New Journal of Physics} \textbf{\bibinfo{volume}{11}},
  \bibinfo{pages}{093029} (\bibinfo{year}{2009}).

\end{thebibliography}

\end{document}